\documentclass{pasj00}     
\SetRunningHead{Hillier et al.}{XXXX}
\Received{XXXX}
\Accepted{XXXX}
\Published{XXXX}         
\begin{document}
\title{Evolution of the Kippenhahn-Schlueter Prominence Model Magnetic Field Under Cowling Resistivity} 
\author{Andrew \textsc{Hillier} and Kazunari \textsc{Shibata}}
\affil{Kwasan and Hida Observatories, Kyoto University, Yamashina-ku, Kyoto, 607-8417, Japan}
\email{andrew@kwasan.kyoto-u.ac.jp}
\and
\author{Hiroaki \textsc{Isobe}}
\affil{Unit of Synergetic Studies for Space, Kyoto University, Sakyo-ku, Kyoto 606-8502, Japan}
\maketitle

\begin{abstract}
We present the results from 1.5D diffusion simulations of the Kippenhahn-Schlueter prominence model magnetic field evolution under the influence of the ambipolar terms of Cowling resistivity. 
We show that initially the evolution is determined by the ratio of the horizontal and vertical magnetic fields, which gives current sheet thinning (thickening) when this ratio is large (small) and a marginal case where a new characteristic current sheet length scale is formed.
After a timespan greater than the Cowling resistivity time, the current sheet thickens as a power law of $t$ independent of the ratio of the field strengths.
These results imply that when Cowling resistivity is included in the model, the tearing instability time scale is reduced by more than one order of magnitude when the ratio of the horizontal field to the vertical field is 20\% or less. 
These results imply that, over the course of its lifetime, the structure of the prominence can be significantly altered by Cowling resistivity, and in some cases will allow the tearing instability to occur.
 
\end{abstract}
 
\section{Introduction}

Quiescent prominences are globally highly stable structures that exist in the corona for days or even weeks.
The characteristic density is $10^{-13} gcm^{-3}$ and the characteristic temperature is $\sim 10^4 K$, which are two orders of magnitude larger and smaller, respectively, than that of the surrounding corona (\cite{HIR1986}).
Using this value for the temperature the pressure scale height can be calculated to be $\Lambda \approx 300km$.
Quiescent prominences form over polarity inversion lines between areas of opposite magnetic flux (\cite{TH1995}) and have a magnetic field strength between $3 \sim 30 G$ (\cite{LER1989}) and a gas pressure of $0.06 dyn cm^{-2}$ (\cite{HIR1986}) giving a plasma $\beta \approx 0.1$.
The magnetic Reynolds number of a quiescent prominence is $R_m \approx 10^7 \sim 10^8$.

Despite the global stability of the quiescent prominence, there are a number of observations that highlight that quiescent prominences are locally highly dynamic.
\citet{ENG1981} reported downflows along vertical filaments, interpreted as downflows along magnetic field lines driven by gravity, of the order of a few $kms^{-1}$.
\citet{KUBO1986} reported flows in a dark filament, observed on disk, that propagated predominately downward with a velocity $\sim 6kms^{-1}$. 
Vorticies inside the prominence of approximately $10^5 km \times 10^5 km$ in size with rotation rates $\sim 30 kms^{-1}$ were reported by \citet{LZ1984}.

Recent observations of quiescent prominences provided by Solar Optical Telescope (\cite{TSU2007}) on the Hinode satellite (\cite{KOS2007}) have shown that on an even smaller small scale, quiescent prominences are highly dynamic structures undergoing constant evolution. 
\citet{BERG2008} found dark upflows, of width $\sim 300 km$, rising from the boundary between cavities that formed at the base of the prominence. 
These dark upflows would rise through a height of around $10Mm$, before forming a mushroom cap profile.

From the above, we can see that globally quiescent prominences are extremely stable structures, but on a smaller scale they are highly dynamic and cannot be considered to be in equilibrium.
One of the main problems in prominence physics is understanding how to capture such dynamics in prominence models.
One possible reason for the formation of the localised dynamics inside the globally stable prominence could result from the partial ionisation of the prominence plasma.
It has been shown that the ionisation fraction of quiescent prominences is around $0.3$ (\cite{HIR1986}).
As this is not a fully ionised plasma, there will constantly be neutral flows inside the prominence, that would stop the formation of any equilibrium that was not dynamic. 
This has been studied observationally by \citet{GIL2007}, who found evidence of cross-field diffusion of neutral particles, which can be viewed as one mechanism to explain mass loss from a quiescent prominence.
Such observations imply that neutral dynamics are playing a role in the evolution of quiescent prominences.

The impact of partially ionised plasma manifests itself in a 1-fluid model in the form of Cowling resistivity (\cite{COW1957}; \cite{BRA1965}).
The impact of Cowling resistivity (composed of the resistive and ambipolar terms) on the magnetic field has already been studied in a wide range of situations.
The formation of sharp structures by ambipolar diffusion created by anti-parallel magnetic fields was studied by \citet{BZ1994}.
The inclusion of ambipolar diffusion was found to greatly enhance the growth rate for the tearing instability at points where the sign of the magnetic field reverses (\cite{CHI1998}).
The three dimensional case is more complicated but ambipolar diffusion was shown to decrease the tearing timescale when the Lorentz force works to compress the plasma.
Cowling resistivity has been discussed as a potential driver in forming coronal nonlinear force free fields in the chromosphere from photospheric fields (\cite{ARB2009}).
It was shown that on a time scale $\sim 15~ mins$ the magnetic field would rearrange itself forming a nonlinear force free field.
\citet{SINGH2010} explained how the ambipolar terms are effective at damping the propagation of Alfv\'en-like wave modes.

This study looks at how the prominence magnetic field, based upon the field used in the Kippenhahn-Schlueter (K-S) prominence model (\cite{KS1957}) evolves under the influence of Cowling resistivity.
The K-S prominence model is based on the simple idea of supporting plasma against gravity through the Lorentz force via magnetic tension. 
This model has been shown to be linearly stable for ideal MHD perturbations (\cite{KS1957}; \cite{ANZ1969}).
\citet{LOPE2005} developed a modified K-S prominence model, created by using the K-S model to describe one sheet inside the prominence and then using an array of sheets with varying properties to create an entire prominence.
This prominence model produced constant flows associated with a structure in local equilibrium but globally not in equilibrium that matched well with observations of flows in prominences.
\citet{PELO2005} discuss the effect of a constant diffusion on the modified K-S model and showed how reconnection could drive a net upward movement of magnetic field associated with a net downward flow of mass. 

\begin{figure*}[t]
\centering
\includegraphics[height=6.5cm]{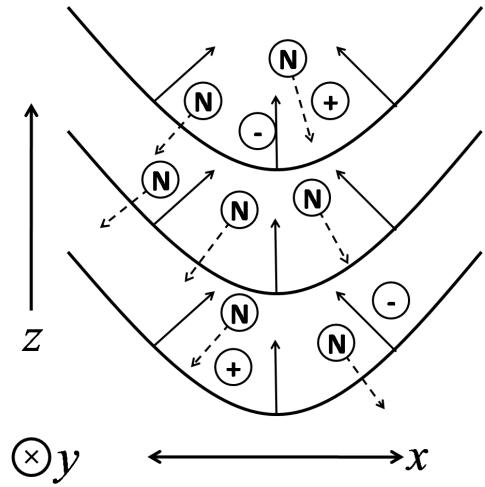}
\caption{K-S magnetic field with neutral and charged particles. Solid arrows show Lorentz force and dashed arrows show direction of neutral motion.}
\label{init}
\end{figure*}

The inclusion of Cowling resistivity and how this effects the current sheet profile of the K-S model are yet to be investigated. 
As the ratio of the Cowling resistivity to the Spitzer resistivity in quiescent prominences is $\eta_C/\eta \approx 1000$ (i.e. the ambipolar terms dominate), the timescale of the deformation of the magnetic field would be greatly reduced through the inclusion of neutral dynamics.
Figure \ref{init} shows the K-S prominence model magnetic field configuration with a fluid made up of neutral and charged particles.
As the neutrals do not directly feel the magnetic field, they will be able to flow out of the current sheet across the magnetic field (dashed arrows show direction of motion), as the is no longer an equilibrium the charged particles are advected by the Lorentz force (solid arrows).
The evolution of  the magnetic field of the K-S model under Cowling resistivity will provide an insight into the nature of prominences and whether it is possible for them to form an equilibrium.
This can then be used to shed some light on how to connect the large scale stability created in prominence models with the local dynamism observed in prominences.

In \S 2 we present the numerical method and in \S 3 we present the results and use these to perform an estimate of the tearing timescale in the K-S model. In \S 4 we present a summary and a discussion outlining the physical meaning behind the results and their implications for quiescent prominences. 

\section{Numerical Method}

The equations for the magnetic field in the K-S prominence model are as follows:
\begin{eqnarray}\label{eqn}
B_{x}(x)=B_{x0} \\
B_{z}(x)=B_{z\infty} tanh \left(\frac{B_{z\infty}}{2B_{x0}}\frac{x}{\Lambda}\right)
\end{eqnarray}
where the x direction is across the prominence ($x=0$ is the centre of the prominence), the y direction (the direction of the current) is along the prominence and the z direction is the vertical direction, $B_{x0} =B_{x}(0)$, $B_{z\infty} =B_{z}(\infty)$ and $\Lambda$ is the pressure scale height. 
As this study focuses purely on the effect of Cowling resistivity on the magnetic field, so the pressure scale height does not have any physical meaning in this setting.
To address this, we normalised the characteristic length scale of the current sheet to be equal to $L$, i.e. $L={2 B_{x0}\Lambda}/{B_{z\infty} }$.
This gives
\begin{eqnarray}
B_{z}(x)=B_{z\infty} tanh \left(\frac{x}{L} \right) \label{eqn2} .
\end{eqnarray}

The induction equation including Cowling resistivity is as follows (\cite{LA2006}):
\begin{equation}\label{cow}
\frac{\partial \mathbf B}{\partial t}= \nabla \left( \mathbf v \times \mathbf B - \eta \mathbf j _\parallel -\eta_C \mathbf j _\perp  \right)
\end{equation}
where $\eta$ is the Spitzer resistivity and $\eta_C=\eta+ ({\xi_n^2}/{\alpha})\mathbf B^2$, where $\xi_n$ and $\alpha$ denote the neutral fraction and the friction coefficient respectively.
We use a 1.5D (${\partial}/{\partial z} =0$) version of this equation, and taking into account that there is no $\mathbf j_\parallel$ component due to the geometry of the problem and that $B_x$ must remain constant to preserve $\nabla \cdot \mathbf B =0$, equation \ref{cow} becomes
\begin{equation}
\frac{\partial B_{z}}{\partial t}= \frac{\partial }{\partial x} \left( \mathbf v \times \mathbf B  -\eta_C \mathbf j \right)
\end{equation}
This work is only concerned with the effect of the diffusion, so the effect of advection of the magnetic field is neglected.
This assumption can be justified as initially the system is in equilibrium, so for the initial $t=\tau_C$ then the dynamics will be dominated by this term.
This is similar to processes observed in molecular cloud fragmentation under Cowling resistivity (\cite{KUDO2008}).
For any evolution after this time, the results from this paper will only provide a handle on the dynamics of the system.
Also when using typical values for quiescent prominences ${\eta_C}/{ \eta}=1000$, so it is possible to neglect $\eta$.
Therefore, this work only presents the solution to the following equation
\begin{equation}\label{induct}
\frac{\partial B_{z}}{\partial t}= -\frac{\partial }{\partial x} \left( \eta_C \mathbf j \right).
\end{equation}
The Cowling resistivity profile is taken to be $\eta_C=({\xi_n^2}/{\alpha}) \mathbf B^2 =\eta_0 \mathbf B^2$ apart from in section (\ref{late}) where a spatially and temporally variable profile is considered.
The time scale is normalised to the Cowling resistivity timescale $\tau_{\eta_c}=L^2/\eta_C$, the length scale to $L$ and the resistivity by $\eta_0 \mathbf B^2$.

To solve the induction equation, we use a central difference scheme that forms part of a two step Lax-Wendroff scheme based on the scheme presented in \citet{UGAI08}.
The calculations were performed on a grid with $10^4$ grid points, unless otherwise stated, where $\Delta x=0.05$.
The outer boundary is a free boundary and the inner boundary a reflective symmetric boundary, in the appendix the case where the outer boundary is a fixed boundary is discussed.

\section{Results}
\subsection{Current Sheet Evolution}

To analyze the temporal evolution of the magnetic field, we perform a series of simulations on a magnetic field as described by equation \ref{eqn} and \ref{eqn2}, for a range of values of $B_{x}$. 
We use a domain of size $X=[0,50L]$, over a timespan of $\sim 100\tau _{C}$.

\begin{figure*}[ht]
\centering
\includegraphics[height=6cm]{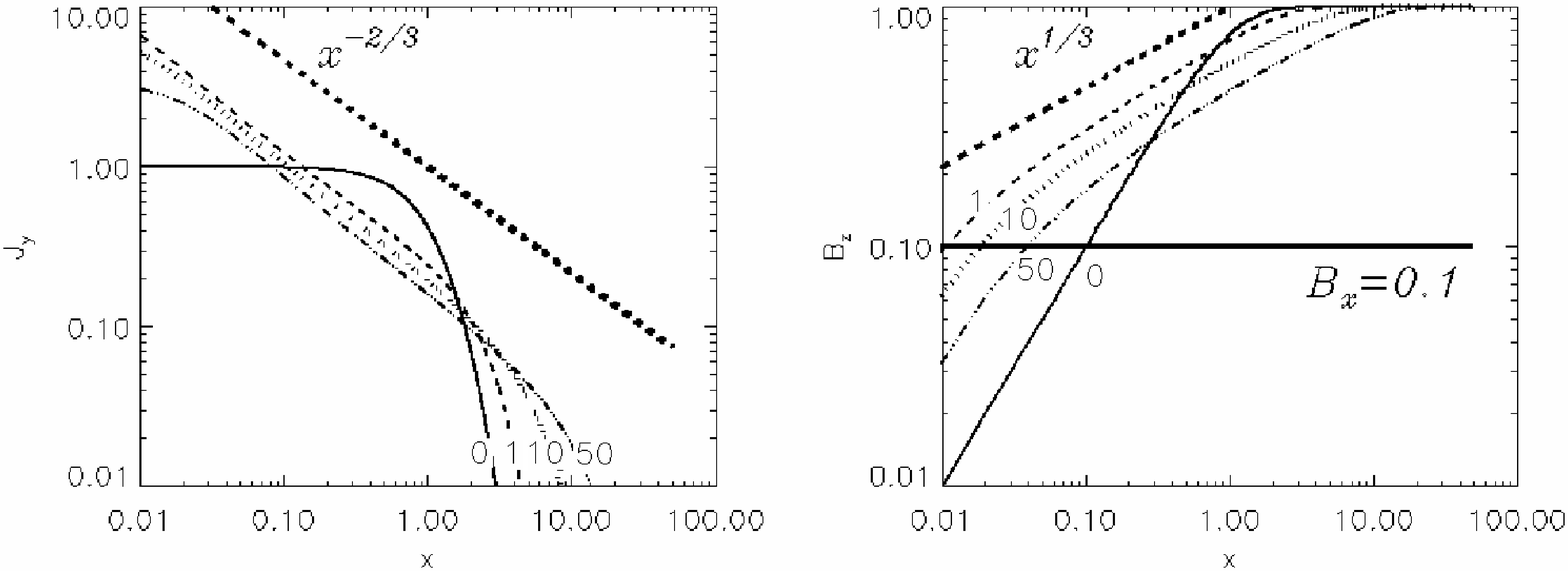}
\includegraphics[height=6cm]{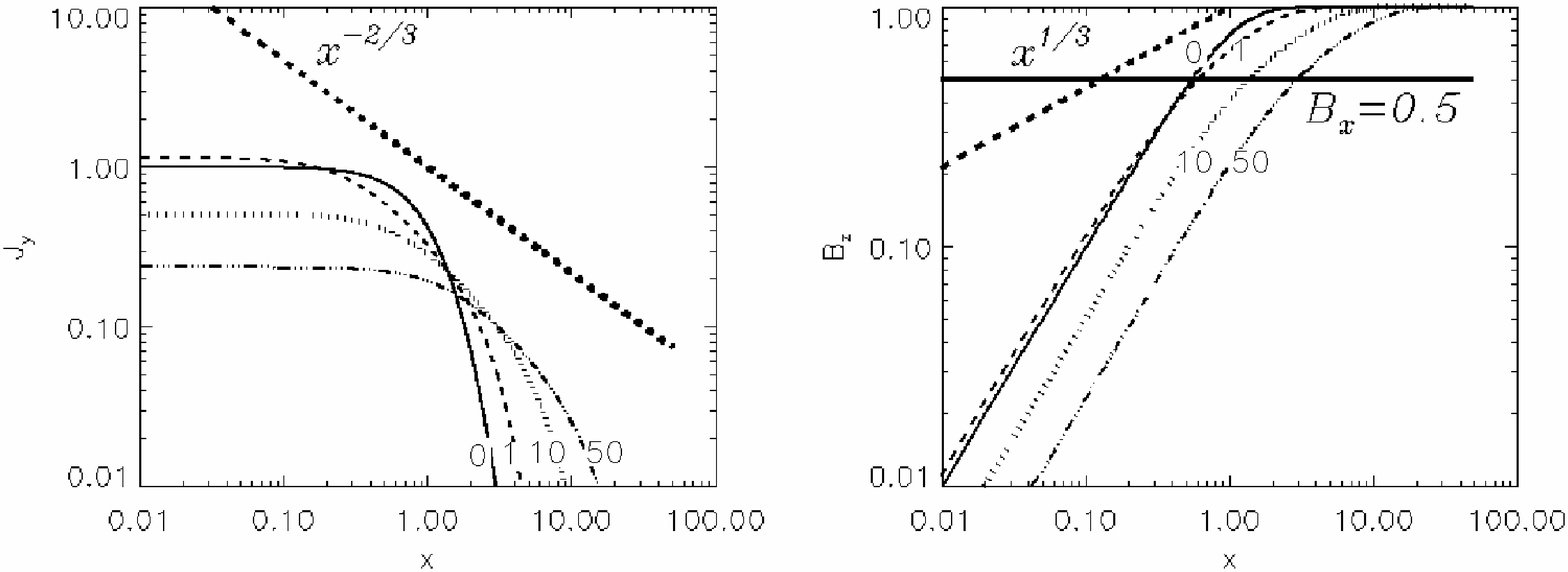}
\caption{Plot of $j_y$ (Left) and $B_z$ (Right) for $B_x=0.5$ (top) and $0.1$ (bottom). The solid, dashed, dotted and three dots + dash lines show the distribution at $t=0,~1,~10$ and $50$}
\label{evol}
\end{figure*}

Figure \ref{evol} shows the evolution of the current sheet and $B_z(x)$ for (top) $B_{x0}/B_{z\infty}=0.1$ and (bottom) $B_{x0}/B_{z\infty}=0.5$.
The solid, dashed, dotted and three dots + dashed lines show the distribution at $t=0,~1,~10$ and $50$ respectively.
The thick dashed line denotes the power $\sim x^{-2/3}$ for the current and $\sim x^{1/3}$ for the magnetic field.
The thick horizontal line shows $B_{x0}/B_{z\infty}$.

It can be clearly seen that the current sheet forms two distinct sections.
The first where $x \in [0,L_{BX}]$, i.e. $B_x>B_z(x)$, and the second where $x \in [L_{BX},\infty]$, i.e. $B_z(x)>B_x$.
The reason for this split in behaviour comes from the definition of the Cowling resistivity, 
\begin{equation}\label{COWRES}
\eta_{C}(x)=(B_x^2+B_z(x)^2)\eta_0.
\end{equation}
As the term $B_x^2 \eta_0$ is a constant, it will work as a uniform resistivity which drives current sheet thickening.
Whereas the  $B_z(x)^2 \eta_0$ term is a variable in $x$, so if the field had no $B_x$ component, this would drive thinning of the current sheet (\cite{BZ1994}).
It can be shown by simple integration that such diffusion  results in the current sheet forming a power law distribution $\propto x^{-2/3}$, and as a result the magnetic field forms a power law distribution where $B_z(x) \propto x^{1/3}$.
As both these processes are working, if $B_{x0}<B_{z\infty}$ there must be some point across the current sheet where they are equal, forming a new characteristic width of the current sheet.

For the case where $B_{x0}/B_{z\infty}=0.1$ (figure \ref{evol} top), current sheet thinning occurs and after $t=\tau_C$ the current sheet has formed the power law distribution described above.
Throughout the evolution of the current sheet, this power law component can be clearly seen at all times.
The same can be said for the magnetic field distribution.
For the case where $B_{x0}/B_{z\infty}=0.5$ (figure \ref{evol} bottom), current sheet thickening and current sheet thinning behaviour can both be observed.
The inner part of the current sheet approximately undergoes constant resistive thickening, maintaining a flat current profile, and the outer part locally appears to form a power law distribution but not to the same extent as the $B_{x0}/B_{z\infty}=0.1$ case.
The current sheet thinning behaviour does not continue for an extended period of time, as shown by the later stages of the evolution in figure \ref{evol} bottom.
Though the dynamics displayed by the magnetic field are not those of a constant resistivity process, no clear power law distribution forms.

We found that the temporal evolution of this new characteristic width, and as a result the magnetic field, can be divided into two separate stages. 
The initial stage as that where approximations of the evolution based on the initial magnetic field distribution still hold.

\subsubsection{Initial Evolution}

\begin{figure*}[t]
\centering
\includegraphics[height=5.5cm]{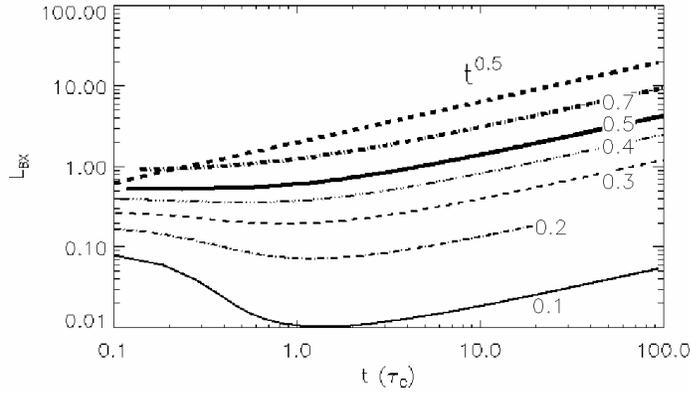}
\caption{Plot of the evolution of the value of $x=L_{BX}$, i.e. the value of x where $B_z(x)=B_x$, for $B_x/B_{z\infty}=0.1$ (solid), $0.2$ (dot + dash), $0.3$ (dashed), $0.4$ (three dots + dash), $0.5$ (thick solid) \& $0.7$(thick dot +dash). The thick dashed line shows $\sim t^{1/2}$}
\label{balance}
\end{figure*}

The initial evolution of the magnetic field is determined by the ratio $B_{x0}/B_{z\infty}$ and the initial current sheet half width $L$. 
Inserting the definition of Cowling resistivity (equation \ref{COWRES}) into the induction equation (equation \ref{induct}), gives
\begin{equation}
\frac{\partial B_{z}}{\partial t}= -\frac{\partial }{\partial x} \left[(B_x^2+B_z(x)^2)\eta_0  \mathbf j \right] .
\end{equation}
Assuming that $\exists x$ s.t. $\partial B_z(x)/\partial t =0$, it is possible to form a comparison of the two components of the diffusion term.
\begin{equation}
\eta _0 B_{x}^2J_{y} \sim \eta _0B_z(x)^2J_{y} 
\end{equation}
This equation can be simplified to give 
\begin{equation}
\frac{B_{x}}{B_{z\infty}} \sim tanh \left(\frac{x}{L} \right)
\end{equation}
from the definition of $B_z$. 
We define the value of $x$ found here ($x=L_{BX}$), as a new characteristic width of the current sheet, where
\begin{equation}\label{WIDTH}
L_{BX}=L tanh ^{-1} \left(\frac{B_{x0}}{B_{z\infty}} \right) \approx L \frac{B_{x0}}{B_{z\infty}},
\end{equation}
it can be seen that this lengthscale is independent of the resistivity profile assumed.
Figure \ref{balance} shows the evolution of $L_{new}$ in time.
It can clearly be seen that this approximation holds for a timespan of $t \approx \tau_{C}$ and for the range $0.3<{B_{x0}}/{B_{z\infty}} <0.7$.

The above range results from comparing the following terms:
\begin{equation}
MAX\left( B_x^2 J_y \right) =MAX\left( B_z(x)^2 J_y \right) 
\end{equation}
As $B_x$ is constant, $MAX\left( B_x^2 J_y \right)=B_x^2 MAX\left( J_y \right)=B_x^2$ giving
\begin{equation}
B_x^2 =MAX\left( B_z(x)^2 J_y \right) \approx 0.25 B_{z\infty}^2
\end{equation}
which gives $B_x/B_{z\infty}\approx 0.5$.
Therefore any marginal case will occur when this relation approximately holds. 
Such evolution is well described by the $B_x/B_{z\infty}=0.5$ case.
As can clearly be seen in figure \ref{evol}, the point at which $B_z(x)=B_x$ initially, the value at $t=\tau_C$ and the linear estimate are approximately the same value.

As stated above, this approximation falls down in the case where ${B_{x}^2}/{B_{z\infty}^2} \ll 1$ and ${B_{x}^2}/{B_{z\infty}^2} \gg 1$.
As the Cowling resistivity is defined as $\eta_{C}=(B_x^2+B_z(x)^2)\eta_0$, when ${B_{x}^2}/{B_{z\infty}^2}\ll1$ then $\eta_{C} \approx B_z(x)^2 \eta_0$.
In this case we find that the linear stage is defined by current sheet thinning as found by \citet{BZ1994}.
This case is well followed by the $B_x/B_{z\infty}=0.1$ case, as shown in figure \ref{evol}.
Though the prediction matches well with the initial value of $x$ where $B_z(x)=B_x$, the value at $t=\tau_C$ is completely different.
The current sheet undergoes thinning, and at $t=\tau_C$ we have a current sheet half width that is $10\%$ of the prediction.

For the case where $1.0 \leq {B_{x0}}/{B_{z\infty}} $ the linear approximation does not hold any more as the new characteristic current sheet half width can only be defined for the case where $B_{x0} \leq B_{z\infty} $, because $\forall x : B_z(x) \leq B_{z\infty}$.
The behaviour of the system can be understood in the same way as used when ${B_{x0}^2}/{B_{z\infty}^2}\ll 1$, if ${B_{z\infty}^2}/{B_{x0}^2}\ll 1$ then $\eta_{C} \approx B_x^2 \eta_0$.
This works in a similar fashion to a constant Spitzer resistivity, which would diffuse the current sheet causing thickening to occur.

\subsubsection{Later Evolution}\label{late}

In the later stage, the presence of the constant horizontal field dominates the dynamics.
It can be seen in figure \ref{balance} that the behaviour of the system changes dramatically from the initial stage. 
In all cases current sheet thickening occurs independent of the value of $B_{x0}/B_{z\infty}$.
The current sheet thickening is driven at a rate $L_{new}\propto t^{1/2}$, which is analogous to current sheet thickening through uniform resistivity applied to a $\delta$ function current sheet.
It has been shown that the half width of a $\delta$ function current sheet's temporal evolution under constant resistivity can be described by $L = 2(\eta_0 t)^{1/2}$ (see, for example, \citet{PR1982}).

\begin{figure*}[t]
\centering
\includegraphics[height=5.5cm]{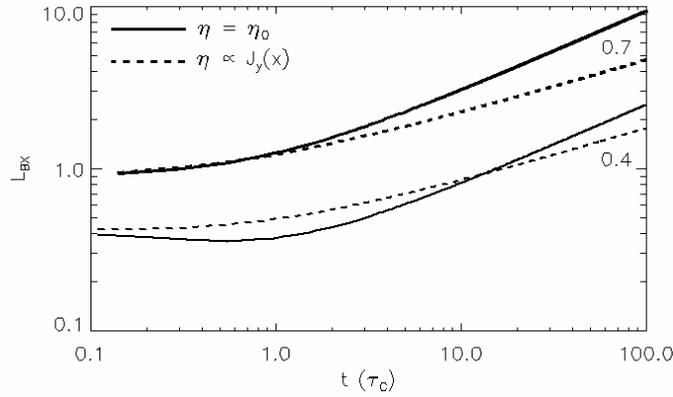}
\caption{Plot of the evolution of the value of $x=L_{BX}$, i.e. the value of x where $B_z(x)=B_x$, for $\eta=\eta_0$ (solid) and $\eta \propto J_y(x)$ (dashed) for $B_x/B_{z\infty}=0.4$ (thin) and $0.7$ (thick)}
\label{varet}
\end{figure*}

The results presented so far have assumed a constant Cowling resistivity.
In a quiescent prominence, the Cowling resistivity would vary  as a function of $x$ as the density and ionisation fraction vary from coronal values to the values at the heart of the prominence.
To create a simple representation of this, under the assumption of constant gravity and hydrostatic equilibrium we have:
\begin{equation}
\rho g =\frac{B_{x0}}{4 \pi}\frac{\partial B_z(x)}{\partial x} \propto J_y(x) .
\end{equation}
From this we can assume that $\eta \propto J_y(x)$ can provide an approximation of the prominence Cowling resistivity profile.
Figure (\ref{varet}) shows the comparison in behaviour between the constant resistivity and the variable resistivity case.
As predicted the initial evolution does not depend greatly on the resistivity profile.
The later evolution, the system takes a power law distribution $\propto t^{0.3}$.
Even with this, it can be seen that up until $t=10 \tau_C$ (approximately the lifetime of a prominence), the current sheet width for both systems is approximately the same.
This allows the results obtained with the constant resistivity to be applied as general results.

A discussion about the effect of the boundary is presented in the appendix.

\subsection{Tearing Instability}

The formula to calculate the growth rate for the tearing instability given by FKR theory (\cite{FKR1962}) is $\omega_{tear} \propto \tau_{A*}^{-1}\alpha^{-2/5}R_{m*}^{-3/5}$, where the $*$ symbol denotes that this is the effective value, i.e. calculated using the current sheet half width as the length scale and $\alpha=ka$.
\citet{SNVH1982} showed numerically that if the constant $\Psi$ approximation is not used the growth rate is $\omega_{tear} \propto \tau_{A*}\alpha^{2/3}R_{m*}^{-1/3}$.
Using the most unstable wavelength, $\alpha_{max}\propto R{m*}^{-1/4} $, the fastest growth rate becomes $\omega_{tear} \propto \tau_{A*}R_{m*}^{-1/2} $.
If we take ${B_{x0}}/{B_{z\infty}}=0.1$ for a current sheet half-width defined as $2{B_{x0}}\Lambda/{B_{z\infty}}$, i.e. that used in the K-S model, and a global magnetic Reynolds number of $R_m=10^7$, we get a tearing time of $\tau_{tear}\propto 10^4s$.
Using the new current sheet half width found in this work, $L_{new}={B_{x0}}L_{old}/{B_{z\infty}}$, the tearing time becomes $\tau_{tear}=300s$.
This result, though, does not take into account the effect of the stabilizing effect of the horizontal field.

The effect of the horizontal field, i.e. a component of the magnetic field the is across the current sheet, was studied in relation to the K-S prominence model by \citet{NISA1982}, where it was found that for values ${B_{x0}}/{B_{z\infty}} \geq 0.1$ the growth rate of the instability scales as $\omega \propto \tau_{A*}^{-1}R_{m*}^{-1}$ which would give a tearing time of the order of $10^7s$ in quiescent prominences.
A similar result was obtained by \citet{HARR1995} in relation to the tearing instability in the Earth's magneto-tail.
Their result implies that as ${B_{x0}}/{B_{z\infty}}$ gets larger, the transition of the growth rate from $\propto R_{m*}^{-3/5}$ to $\propto R_{m*}^{-1}$ happens at smaller values of $R_m*$.
For example, for  ${B_{x0}}/{B_{z\infty}}=10^{-3}$ this transition happens at $R_{m*}=10^5$.
As such weak horizontal fields would be unrealistic in quiescent prominences, it is very unlikely that the tearing instability could occur without an outside process driving it.  
\citet{SANI1983} showed that the tearing instability could be excited in such magnetic fields if driven by an outside force, in their case provided by short wavelengths fast mode waves.
We believe that thinning of the current sheet caused by Cowling resistivity could also allow the tearing instability to occur.

\begin{figure*}[ht]
\centering
\includegraphics[height=5.5cm]{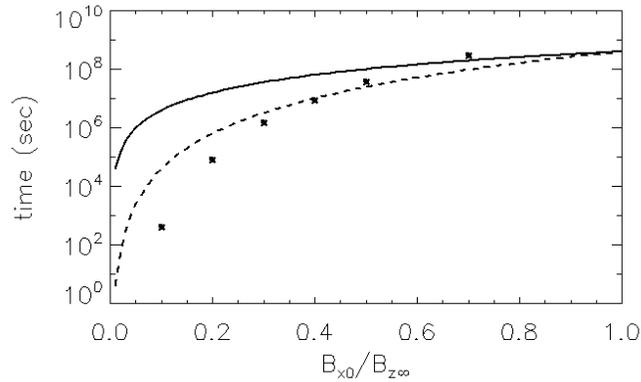}
\caption{Plot of the of tearing time scale against $t$(sec) for different initial values of $B_{x0}/B_{z\infty}$. The solid line is for the original K-S model, the dashed line show the tearing timescale based on the linear estimate of the current sheet width under Cowling resistivity (see equation \ref{WIDTH}) and the stars denote the tearing timescale for the current sheet width found from simulations (see fig. \ref{balance}) at $t=\tau_C$}
\label{timescale}
\end{figure*}

Using the new current sheet widths found, it will be possible to make an estimate for the tearing instability timescale in a quiescent prominence, this estimate is shown in figure \ref{timescale}.
The solid line denotes the estimate for the tearing timescale based on the current sheet half width from the K-S model, the dashed line denotes the tearing timescale based on the linear estimate for the current sheet half width under Cowling resistivity and the stars denote the tearing timescale for the current sheet half width found from the simulation results, these are taken at when $t= \tau_C$.

Figure \ref{timescale} clearly shows that the inclusion of Cowling resistivity reduces the tearing timescale when $B_{x0}/B_{z\infty}< 0.5$.
The Cowling resistivity timescale is approximately $10^5 s$, during which time we would expect the magnetic field to have restructured itself due to the flow of the neutral particles, therefore any tearing time that is longer than this can be viewed as unrealistic. 
From these results, when we include Cowling resistivity, we can say that for the case where $B_x/B_{z\infty}<0.2$, it is possible that tearing could occur.
Whereas, when we do not include the effect of Cowling resistivity, it is only possible where $B_x/B_{z\infty}\ll 1$, that is to say, unrealistically small values for the horizontal field component.
Based on this result, we feel that the inclusion of neutral dynamics could be very important when studying quiescent prominences.
However, when applied to active region prominences, i.e. those with strong horizontal fields, the timescale for the instability is around $10^8s$ so would not be able to occur inside the life time of the prominence, so such dynamics may not be important.

\section{Discussion and Summary}

One of the big problems in understanding quiescent prominence dynamics is explaining the highly dynamic features that are seen in the prominence using a model that has been designed to explain the global stability of the prominence.
Until now, it has been commonplace to ignore the physics that are involved with partially ionized plasmas.
Using values that are typical for quiescent prominences gives timescale for these effects of approximately $1.5$ days, which is shorter than the average lifetime of a quiescent prominence.
Therefore the physical effects from partially ionized plasma could be very important for creating the highly dynamical features seen in quiescent prominences.

In this work we have shown the impact of including the neutral component in a 1-fluid model, by using the addition of the Cowling resistivity to the induction equation, to understand how the magnetic field in the Kippenhahn-Schlueter prominence model behaves.
We have shown that the temporal evolution of the magnetic field, characterized by the evolution of the current sheet, which is defined by the value of $x$ such that $B_z(x)=B_{x0}$
The evolution can be divided into two clear regimes, the initial evolution which can be determined by the ratio $B_{x0}/B_{z\infty}$ and the later evolution where the current sheet thickens with $L_{new} \propto t^{1/2}$.

Without performing full 3D MHD simulations that include the Cowling resistivity, it is impossible to fully describe the dynamics that would be created inside the K-S model, though these relations provide a handle on how the model will evolve.
As the plasma will flow in the opposite direction to the Lorentz force, for $x<L_{B_x}$ the tension force dominates the magnetic pressure in the K-S model.
This implies that the plasma will flow downward.
For $x>L_{B_x}$ the magnetic pressure force dominates the tension force, therefore there will be an outflow occurring along the $x$-direction.

This study has allowed estimates of the tearing time scale to be calculated for the K-S model, giving the ratio $B_x/B_{z\infty}$ at which tearing is possible.
It is clear that the inclusion of Cowling resistivity radically reduces the tearing time scale in the K-S model.
As a result, the range of values of $B_x/B_{z\infty}$ at which we are likely to have tearing occur increases from $0 \sim 0.01$ to $0 \sim 0.2$.
This implies that any numerical simulations of the tearing instability in solar prominences should include Cowling resistivity. 
This result, when applied to solar prominences, could give a handle on the magnetic field strength at positions where plasmoid ejection is observed.

It is very important to extend this study to include fluid effects by solving the full MHD equations.
The results in this study for the initial evolution of the magnetic field ($0<t \leq 1$) should match well with any results from MHD simulations, but after that the fluid effects should become important.
This is because $\tau_C/\tau_A \approx 10000$.
Therefore, once the flows have been initiated by the decoupling of the magnetic field from the plasma via the Cowling resistivity, these flows could have a significant effect on the results.
In this work, we found that the later evolution of the magnetic field for $\eta_C=\eta_0$ was defined by a current sheet thickening with $L_{new} \propto t^{1/2}$, without the inclusion of fluid flow this can only be viewed as a handle on the potential dynamics of the system.
In reality, 2-fluid simulations with ionisation from coronal x-rays would be necessary to truly understand the dynamics of the system.
How this is effected by fluid flow would be a very interesting future research topic.

The authors would like to thank the staff and students of Kwasan and Hida observatories for their support and comments.
This work was supported in part by the Grant-in-Aid for the Global COE program ``The Next Generation of Physics, Spun from Universality and Emergence'' from the Ministry of Education, Culture, Sports, Science and Technology (MEXT) of Japan.
AH is supported in part by the Japanese government (Monbukagakusho) scholarship from the Ministry of Education, Culture, Sports, Science and Technology (MEXT) of Japan. HI is supported by the Grant-in-Aid for Young Scientists (B, 22740121).
 
 \appendix
 
\section*{Boundary Effects}\label{Bound}

In the main body of this work, the temporal evolution of the current sheet was discussed.
However, for numerical simulations it is important to know the effect of the boundary on the system.
As stated previously, a free boundary condition (${\partial B_z}/{\partial x}=0$) was used for the calculations presented in this work.
We can simply calculate the steady state formed with this boundary condition by rewriting equation \ref{cow} and neglecting the Spitzer resistivity. 
We can write
\begin{equation}
\frac{\partial \mathbf B}{\partial t}=\nabla \times [(\mathbf v \times \mathbf v_C)\times \mathbf B]
\end{equation}
where $\mathbf v_C=({\xi^2_n}/{\alpha})\mathbf j \times \mathbf B$.
Therefore the effect of Cowling resistivity can be understood to be equivalent to advecting the magnetic field in the direction of the $\mathbf j \times \mathbf B$ vector, which in terms of the equation of motion is the Lorentz force. 
Therefore, in general, any steady state of the magnetic field will be the state where $\mathbf j \times \mathbf B =0$, if the boundary allows such a state to be formed. 

The free boundary condition used in this study allows the magnetic field to relax to a $\mathbf j \times \mathbf B=0$ state, i.e. $\forall x: B_z(x)=0$ leaving a constant horizontal field.
A symmetric boundary would allow the same process to occur.
A fixed boundary (${\partial B_z(L_{BND})}/{\partial t}=0$ where $L_{BND}$ is the value of $x$ at the boundary) would not allow such a state to be reached, because the angle of the magnetic field at the boundary is fixed.
Therefore, any steady state that is reached will be such that $\exists x$ such that $\mathbf j \times \mathbf B \neq 0$.
In this subsection, we will investigate the steady state formed when using a fixed boundary.

\begin{figure*}[ht]
\centering
\includegraphics[height=5.5cm]{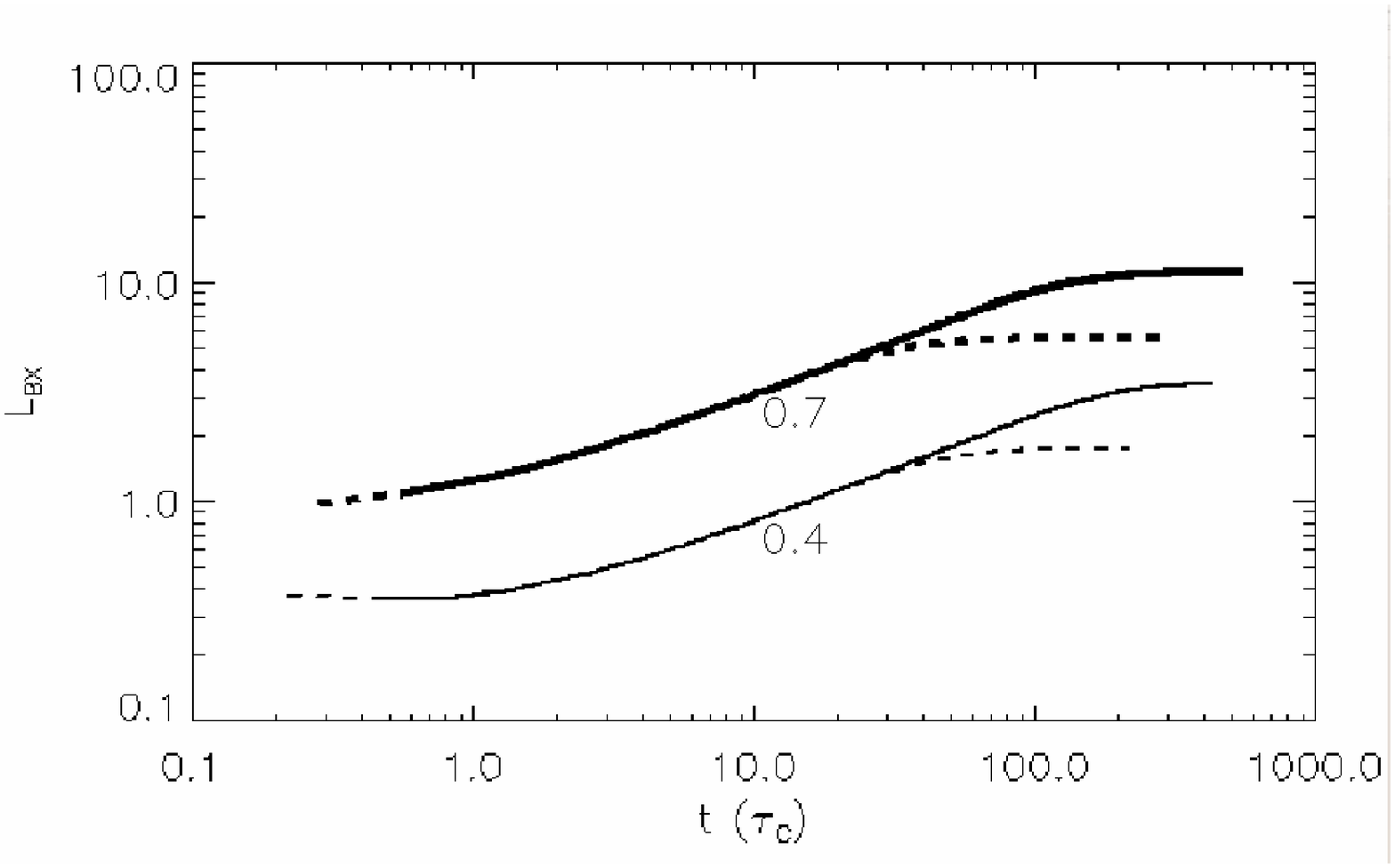}
\caption{Plot of the evolution of the value of $x=L_{BX}$, i.e. the value of $x$ where $B_z(x)=B_x$, for $B_x/B_{z\infty}=0.4$ (dashed-$L_{BND}=10L$ and solid-$L_{BND}=20L$) and $0.7$ (thick dashed-$L_{BND}=10L$ and thick solid-$L_{BND}=20L$)}
\label{balancesteady}
\end{figure*}

Figure \ref{balancesteady} shows the temporal evolution of $L_{BX}$ for a solid boundary ($L_{BND}$) at $x=10L$ \& $20L$.
The dashed and solid lines correspond to the case where $B_x/B_{z\infty}=0.4$ for $L_{BND}=10L$ and $20L$ respectively, and the thick dashed and thick solid lines correspond to the case where $B_x/B_{z\infty}=0.7$ for $L_{BND}=10L$ and $20L$ respectively.
It is clear that once the steady state has been reached a characteristic current sheet half width is formed that it has a dependence on the value of $L_{BND}$ and $B_x^2/B_{z\infty}^2$.
Analysis of the results shows that the current sheet half width of the steady state $L_{steady} \propto L_{BND} {B_x^2}/{B_{z\infty}^2}$.

As shown in section \ref{Bound}, a steady state can be formed whilst the field lines remain curved.
This is only applicable for a fixed boundary, which does not apply to Solar prominences as the magnetic field as the surrounding corona would act like a free boundary due to its low density.
This type of magnetic field equilibrium could be applicable to the emergence of flux tubes or sheets that rise from below the photosphere into the solar atmosphere, where the photosphere can act as a fixed boundary.
The small ionisation fraction in the chromosphere means that if the emergence of the flux tube is sufficiently slow, the structure of the magnetic field could be greatly altered to form an equilibrium.

\end{document}